\documentclass[conference]{IEEEtran}
\IEEEoverridecommandlockouts
\usepackage{cite}
\usepackage{hyperref}
\hypersetup{
	colorlinks=true,
	linkcolor=blue,
	filecolor=magenta,      
	urlcolor=cyan,
	citecolor=red
}
\usepackage{enumitem}  
\usepackage{physics}
\usepackage{amsmath,amssymb,amsfonts}
\usepackage{algorithmic}
\usepackage{graphicx}
\usepackage{textcomp}
\usepackage{varwidth}
\usepackage{xcolor}
\def\BibTeX{{\rm B\kern-.05em{\sc i\kern-.025em b}\kern-.08em
    T\kern-.1667em\lower.7ex\hbox{E}\kern-.125emX}}

\begin{document}

\title{
Optimal Entanglement Distillation Policies\\ for Quantum Switches
\thanks{KPS acknowledges support from NSF grant no. 2134891 through sub-contract from University of Maryland. E-mail:vik80@pitt.edu, nkc16@pitt.edu}
}

\author{\IEEEauthorblockN{Vivek Kumar, Nitish K. Chandra, \& Kaushik P. Seshadreesan}
\IEEEauthorblockA{\textit{Department of Informatics \& Networked Systems}\\
\textit{School of Computing \& Information, University of Pittsburgh}\\
Pittsburgh, PA 15213, USA}
\and
\IEEEauthorblockN{Alan Scheller-Wolf \& Sridhar Tayur}
\IEEEauthorblockA{
\textit{Tepper School of Business} \\
\textit{Carnegie Mellon University}\\
Pittsburgh, PA 15213, USA}
}

\maketitle

\begin{abstract}
The quantum switch is a vital component for the development of the quantum internet. 
In an entanglement distribution network, the function of a quantum switch is to generate {\em elementary} entanglement with its clients followed by entanglement swapping to distribute {\em end-to-end} entanglement of {\em sufficiently high fidelity} between clients. 
The threshold on entanglement fidelity is any quality of service requirement specified by the clients as dictated by the application they run on the network. 
We consider a discrete time model for a quantum switch that attempts generation of fresh elementary entanglement with clients in each time step in the form of maximally entangled qubit pairs, or {\em Bell pairs}, which succeed probabilistically; the successfully generated Bell pairs are stored in noisy quantum memories until they can be swapped. 
We focus on establishing the value of entanglement distillation of the stored Bell pairs prior to entanglement swapping in presence of their inevitable aging, i.e., decoherence: For a simple instance of a switch with two clients, an exponential decay of entanglement fidelity, and a well-known probabilistic but heralded two-to-one distillation protocol, given a threshold end-to-end entanglement fidelity, we use the Markov Decision Processes framework to identify the optimal action policy - to wait, to distill, or to swap that maximizes throughput. 
We compare the switch's performance under the optimal distillation-enabled policy with that excluding distillation. 
Simulations of the two policies demonstrate the improvements that are possible in principle via optimal use of distillation with respect to average throughput, average fidelity and jitter of end-to-end entanglement, as functions of fidelity threshold. 
Our model thus helps capture the role of entanglement distillation in mitigating the effects of decoherence in a quantum switch in an entanglement distribution network, adding to the growing literature on quantum switches.
\end{abstract}

% Note that keywords are not normally used for peerreview papers.

% For peer review papers, you can put extra information on the cover
% page as needed:
% \ifCLASSOPTIONpeerreview
% \begin{center} \bfseries EDICS Category: 3-BBND \end{center}
% \fi
%
% For peerreview papers, this IEEEtran command inserts a page break and
% creates the second title. It will be ignored for other modes.
\IEEEpeerreviewmaketitle

\section{Introduction}
% The very first letter is a 2 line initial drop letter followed
% by the rest of the first word in caps.
% 
% form to use if the first word consists of a single letter:
% \IEEEPARstart{A}{demo} file is ....
% 
% form to use if you need the single drop letter followed by
% normal text (unknown if ever used by the IEEE):
% \IEEEPARstart{A}{}demo file is ....
% 
% Some journals put the first two words in caps:
% \IEEEPARstart{T}{his demo} file is ....
% 
% Here we have the typical use of a "T" for an initial drop letter
% and "HIS" in caps to complete the first word.
The future quantum internet~\cite{Wehner2018-rx}---a global-scale quantum network of networks of interconnected quantum computers, sensors and other devices, lies at the pinnacle of the second quantum revolution~\cite{Dowling2003-by} that is currently underway. 
It will coexist alongside the present day classical internet, augmenting its security by way of quantum cryptography~\cite{Pirandola2020-tt} protocols such as quantum key distribution~\cite{Mehic2020-rb}, while also enabling distributed quantum information processing~\cite{Vardoyan2022-gq, Van_Meter2016-dd} and secure delegated quantum computation~\cite{Kashefi2017-et, Fitzsimons2017-zk} in the cloud that cannot be supported solely by today's classical internet. 
The main challenge in realizing entanglement-distribution-based quantum networks as part of the quantum internet lies in reliably creating quantum entanglement across large distances at high rates, because quantum signals deteriorate exponentially with the transmission distance~\cite{Pirandola2017-sc, Azuma2016-bm, Takeoka2014-zp}. 
Quantum repeaters~\cite{Muralidharan2016-xj, Munro2015-pa, Guha2015-lk} help overcome this challenge; when introduced in a quantum network they boost quantum signals using quantum sources, memories, detectors and logic, thereby enabling distribution of higher quality entanglement at enhanced rates. 

Apart from quantum repeaters, also essential for the development of quantum networks are quantum switches. 
A quantum switch is firstly also a quantum repeater in itself that enhances the rate of reliable quantum transmissions that pass through it, but additionally due to its ability to handle quantum communications between multiple users, can direct quantum transmission flows along different paths in the network, and thereby enable large-scale quantum networks of arbitrary topologies~\cite{Van_Meter2014-tq}. 
A quantum switch in an entanglement distribution network works by first establishing entanglement individually with each of its {\it clients}, i.e., nodes that share direct optical links with it. 
We refer to these as {\em link-level} entanglement or {\em elementary} entanglement. 
Fresh link-level entanglement is typically modeled as a pair of maximally entangled qubits, or {\it Bell pairs}, where the Bell pairs are generated at constant rates that factor in the rate of attempted link-level entanglement generation and success probability of attempts. 
The switch then distributes {\it end-to-end} entanglement between any subset of clients by connecting appropriate link-level entanglement via entanglement swapping. 
The measurements involved in entanglement swapping are most generally probabilistic, which when successful, lead to end-to-end entanglement between the clients. 
The decisions of the switch are governed by a switching policy that is designed to optimize any chosen metric, while catering to quality of service requirements as specified by its clients such as, e.g., a minimum threshold fidelity of end-to-end entanglement. 

Different models for the quantum switch in an entanglement distribution network and policies thereof have been studied. 
This includes, e.g., a switch that only serves bipartite entanglement~\cite{vardoyan2021exact,vardoyan2019stochastic} between different pairs of its clients by performing Bell-state measurements (BSM), a switch that serves multipartite entanglement between different subsets of clients using Greenberger-Horne-Zeilinger (GHZ) basis measurements~\cite{Nielsen2002-ht} on link-level entanglement~\cite{Nain2020-ed}, and a switch that does both~\cite{vardoyan2023}.
Some of these works consider the effects of decoherence on link-level entanglement modeled by finite lifetime for quantum memories~\cite{vardoyan2021exact,vardoyan2019stochastic}. 
While some of the models consider static, switch-centric metrics such as the sum throughput, and are agnostic to client demands~\cite{vardoyan2023,Vardoyan2022-gq,vardoyan2021exact,vardoyan2019stochastic}, some take into account client requests for the different types of end-to-end entanglement and try to maximize request service rates by suitably prioritizing requests such that queues for the different request types (that indicate backlogs) do not grow unboundedly~\cite{dai2022capacity,Vasantam2021-qu}. 
A quantum switch model for continuous variable quantum encodings (instead of single-photon-based qubit encodings) that in addition to request rates and stability of request queues also takes into account polarities of elementary entanglement for optimally enhanced end-to-end entanglement rates has been recently explored~\cite{tillman2022continuous, Tillman2022-pq,Seshadreesan2020-pl}. 
%Recent studies have focused on developing protocols for scheduling entanglement swapping in quantum switches. 
%In Ref., the impact of entanglement requests received randomly on the switch's ability to support stable rate vectors was analyzed. 
Most recently, for a switch model where multiple, noisy, imperfect link-level Bell pairs are generated per client that live only one time step (they are discarded at the end of each time step unless used), the use of nested two-to-one probabilistic but heralded entanglement distillation~\cite{BBPSSW96,DBCZ99} to jump over a fidelity threshold on end-to-end entanglement has been studied~\cite{panigrahy2022capacity}. 
The work presented a performance comparison for distilling elementary entanglement followed by entanglement swap versus entanglement swap followed by end-to-end entanglement distillation.

%It has been known that Markov decision processes (MDPs) \cite{Puterman2014-fz} can facilitate a well-defined approach to mathematically model protocols for quantum networks \cite{khatri2022design}. 

In this work, we consider a model for a switch in an entanglement distribution network for the simplest instance of the switch with just two clients and bipartite qubit entanglement distribution. Our focus is on establishing the value of distillation in mitigating the effects of decoherence: We model link-level Bell pairs with clients as being retained over multiple time steps in the presence of decoherence with aging as they wait to be paired with link-level counterparts. While link-level Bell pairs wait to be swapped, we allow entanglement distillation between noisy Bell pairs across each link as a possible action to maintain the quality of the links for the switch. 
% Our primary goal is to understand the value of distillation on the performance of the quantum switch. 
Specifically, we utilize a Markov Decision Process (MDP) framework within a discrete-time model with: (a) probabilistic creation (arrivals) of link-level entanglement on either side of the switch; (b) finite buffers to store links which decohere over time as they await entanglement swaps; and 
% having to wait (in case there is no Bell pair to swap with on the other client). 
c) the option to perform
% to entanglement swapping, the switch is allowed to perform 
two-to-one entanglement distillation between two Bell pairs on the same link to improve the fidelity of the resulting distilled Bell pair if and when successful. 
Given that this operation is probabilistic in its success (both links can be lost with non-zero probability), one has cautiously choose when to do it, ensuring the benefit is worth the risk. The MDP outputs the optimal action policy to balance the benefits and risks. We use simulation to compare the performance of the switch with another that does not allow any distillation
% also based on the MDP framework by performing simulations 
to evaluate average throughput and jitter of end-to-end entanglement generation as function of a target entanglement fidelity threshold.

% You must have at least 2 lines in the paragraph with the drop letter
% (should never be an issue)
%I wish you the best of success.

% needed in second column of first page if using \IEEEpubid
%\IEEEpubidadjcol

The paper is organized as follows. 
In Sec.~\ref{prelim}, we include some preliminaries on quantum entanglement and its manipulation. 
In Sec.~\ref{qswitch}, we outline our model for the quantum switch. In Sec.~\ref{QS_MDP}, we discuss the Markov Decision Process and use it to formulate quantum operations for the quantum switch. In Sec.~\ref{Results}, 
 we describe the details of the optimal policy and examine the simulation results. The concluding section (Sec.~\ref{Future_Direc}) summarizes our findings and outlines future research directions.

\section{Preliminaries \label{prelim}}
Here, we briefly review the quantum operations relevant to the quantum switch. We consider finite dimensional quantum systems, i.e., systems whose state spaces are finite dimensional Hilbert spaces. 
We denote them by capital letters $A$, $B$, $C$, etc. 
States of a quantum system are most generally described by positive semi-definite, trace one operators, known as density operators. 
Rank-1 density operators, i.e., unit-norm vectors, represent pure states.
%Let $\mathcal{S(H)}$ denote the set of density operators defined on a finite, $d$, dimensional Hilbert space $\mathcal{H}$.  

For the composite quantum system comprised of two qubits $A$ and $B$, i.e., whose associated Hilbert space is the tensor product space $\mathcal{H}_2\otimes \mathcal{H}_2$, a set of orthogonal basis states is given by Bell states:
\begin{align}
|\Phi^\pm\rangle_{AB}=\left(|0\rangle_A\otimes |0\rangle_B\pm|1\rangle_A\otimes |1\rangle_B\right)/\sqrt{2},\nonumber\\
|\Psi^\pm\rangle_{AB}=\left(|0\rangle_A\otimes |1\rangle_B\pm|1\rangle_A\otimes |0\rangle_B\right)/\sqrt{2}.
\end{align}
Werner states of the two qubit system are density operators of the form:
% \begin{align}
% \rho_{AB} = F|\Phi^+\rangle\langle\Phi^+|_{AB}+\frac{1-F}{3}\left(|\Phi^-\rangle\langle\Phi^-|_{AB}+
% \\
% |\Psi^+\rangle\langle\Psi^+|_{AB}+|\Psi^-\rangle\langle\Psi^-|_{AB}\right).
% \end{align} 
\begin{align}
\rho_{AB} =& F|\Phi^+\rangle\langle\Phi^+|_{AB}+\frac{1-F}{3}(|\Phi^-\rangle\langle\Phi^-|_{AB}\nonumber
\\
&+|\Psi^+\rangle\langle\Psi^+|_{AB}+|\Psi^-\rangle\langle\Psi^-|_{AB}).
\end{align} 
The parameter $F=\operatorname{Tr}\left(\rho_{AB}|\Phi^+\rangle\langle\Phi^+|_{AB}\right)$, $1\geq F\geq 0$, captures the fidelity of the Werner state with the target Bell state $|\Phi^+\rangle$. 
Alternatively, the Werner state can also be written as
\begin{equation} 
\rho = x|\Phi^{+}\rangle\langle\Phi^{+}|+\frac{1-x}{4}\mathbb{I},
\end{equation}
where $x=(4F-1)/3$ and $\mathbb{I}$ is a $4\times4$ identity operator on $\mathcal{H}_2\otimes \mathcal{H}_2$. 

\subsection{Entanglement Swapping}
When two clients of a quantum switch share two-qubit entangled states $\rho_{AC}$ and $\rho_{BD}$, respectively, with the switch, where qubits $A$ and $B$ are at the clients and qubits $C$ and $D$ are at the switch, by performing a Bell state measurement on qubits $C$ and $D$, the quantum switch creates an entangled state between qubits A and B, leading to end-to-end entanglement between the users. 
This phenomenon is known as entanglement swapping, where qubits A and B become correlated with each other, despite not having directly interacted before. 

Let $\rho_{AC}$ and $\rho_{BD}$ be Werner states with entanglement fidelity $F_{1}$ and $F_{2}$ (with respect to the $|\Phi^{+}\rangle$ Bell state), respectively,
\begin{align}
    \rho_{AC} = x|\Phi^{+}\rangle\langle\Phi^{+}|+\frac{1-x}{4}\mathbb{I}\nonumber,\\
    \rho_{BD} = y|\Phi^{+}\rangle\langle\Phi^{+}|+\frac{1-y}{4}\mathbb{I},
\end{align}
where, $x=(4F_{1}-1)/3, y=(4F_{2}-1)/3\in \mathbb{R}$.
The parameters $x$ and $y$ determine whether or not the respective Werner states are entangled or separable. 
For values $1/3 < \{x, y\} \leq 1$, the states are entangled which translate to entanglement fidelity values $F\geq \frac{1}{2}$~\cite{Azuma2006-dh}.
When the qubits $C$ and $D$ in $\rho_{AC}\otimes\rho_{BD}$ are projected into the Bell basis, we get an output Werner state $\rho_{AB}$ with fidelity~\cite{SenDe2005-eq},
\begin{equation}
    F_{12} = \frac{3xy+1}{4},
\end{equation}
or equivalently
\begin{equation}
     F_{12} = \frac{((4F_{1}-1)(4F_{2}-1)/3)+1}{4}.
     \label{Werner_swap}
\end{equation}

\subsection{Entanglement Distillation}
Entanglement distillation refers to the process of transforming multiple noisy imperfect entangled states into fewer higher quality entangled states. 
For a pair of Werner states $\rho_{A_1B_1}$ and $\rho_{A_2B_2}$ with fidelities $F_1$ and $F_2$, a well-known entanglement distillation protocol was given by Bennett et al.~\cite{BBPSSW96}. 
It is a probabilistic but heralded protocol that involves applying CNOT operations with $A_1$ and $B_1$ as controls and $A_2$ and $B_2$ as targets, followed by measuring $A_2$ and $B_2$ in the computational basis. 
The protocol succeeds when both measurement yield identical outcomes, and fails when the outcomes are different. 
The success probability is given by
\begin{align}
p_\textrm{succ} = \frac{8}{9}F_1F_2-\frac{2}{9}(F_1+F_2)+\frac{5}{9}.
\label{Werner_distill_p}
\end{align}
When successful, the output of the protocol is a Werner state of fidelity given by
\begin{align}
F_{12}=\frac{1}{p_\textrm{succ}}\left(\frac{10}{9}F_1F_2-\frac{1}{9}(F_1+F_2)+\frac{1}{9}\right).
\label{Werner_distill_F}
\end{align}
Entanglement distillation can thus help surpass quality of service requirements in terms of threshold entanglement fidelities in a quantum network at the expense of many (typically lower-fidelity) link-level entanglements. In the case of probabilistic protocols, these also allow a departure from determinism.
%If one considers a CNOT gate infidelity $\epsilon_G$ and measurement error probability $\xi$, the expression for $F_{12}$ is calculated in Ref. \cite{dur1999quantum}. %byDur et al.~\cite{DBCZ99}
\section{Quantum Switch \label{qswitch}}

\begin{figure}[htp]
       \centering
 \includegraphics[width=9cm]{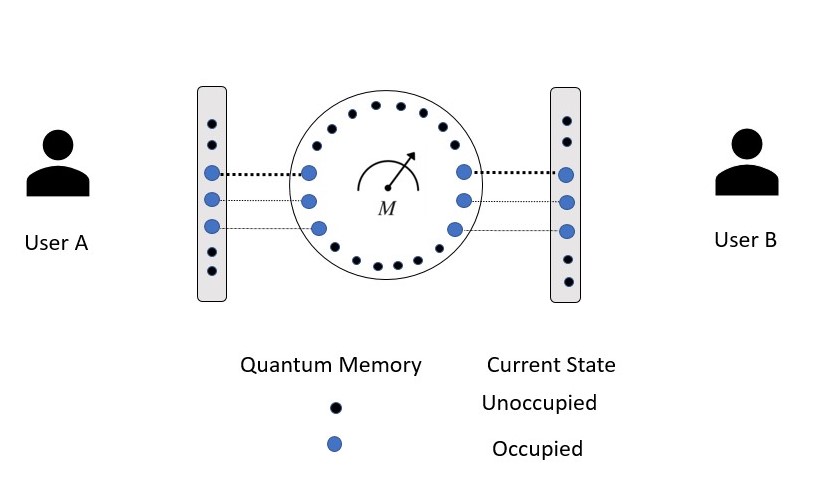}
        \caption{A schematic of quantum switch with three users (A, B) where thicker dotted line represents a newer elementary link.}
        \label{fig:schematic}
    \end{figure}

We consider the simplest case of a quantum switch with just two clients, as illustrated in the schematic in Fig.~\ref{fig:schematic}. 
The clients periodically attempt to establish a single link-level entanglement in the form of Bell pairs with the switch at a pre-defined rate $R_{\textrm{clock}} = \frac{1}{T_{\textrm{clock}}}$ per unit time step of the switch, where $T_{\textrm{clock}}$ is the duration of a time step (that can be set to 1 without loss of generality), and the attempts succeed with probabilities $\lambda_1, \lambda_2$, respectively. 
%If the entanglement generations are successful, a two-qubit Bell state is shared where one qubit is with the switch and the other with the user. 
The function of the switch is to connect successfully generated link-level Bell pairs with the two clients. 
However, it can do so only when: i) they are available with both clients, and ii) the resulting state after entanglement swapping at the switch can yield a Bell pair of entanglement fidelity higher than a target threshold $F_{\textrm{th}}$. 
Until then, Bell pairs that are successfully generated with any one of the two clients alone are stored in quantum memory. 
The link-level Bell pairs are modeled as Werner states. 
When stored in quantum memory, the Werner state fidelity is modeled to decay exponentially with time (starting from $F=1$ at time $t=0$). 
Additionally, the link-level Bell pairs are discarded as unusable after a certain, finite number of time steps $m^*$, i.e., time $t=m^*$ (when the fidelity reaches a cutoff fidelity $F^*$). 
In other words, for a link-level Bell pair of age $m\in (0,1,2....,m^{*})$, $F(m) =e^{-\alpha m}$, where $\alpha$ is a decoherence parameter. 
Since fidelity $F<1/2$, would imply a separable state, a possible value for the fidelity cutoff is $F^*=1/2$. 
However, more generally, for any cutoff fidelity $F^{*}$ of a link-level Bell pair, the value of the decoherence parameter, $\alpha = -\frac{\ln (F^{*})}{m^{*}}$.

The switch functions as follows: Given one or many successfully generated link-level Bell pairs among the two clients, if there are pairs on both sides,
% when a link-level Bell pair also arrives on the second client, 
the switch can either connect any pair from one client directly with any one of the Bell pairs of the other client that are stored and active, or when equipped with the option of entanglement distillation, perform probabilistic entanglement distillation between any pair of stored link-level Bell pairs. We assume at the start of each period the new pairs arrive (if successful) after this up to one action (swap/distill) can be performed.

% connect it to a link-level Bell pair that is the output of the
% probabilistic entanglement distillation.
% procedure that is performed

%As soon as the switch has a sufficient number of Bell pairs connected, it performs an entangling measurement to create end-to-end entanglement between the two users according to the switching policy.

%Let the age of a particular link-level Bell pair in storage be $m\in (0,1,2....,m^{*})$ where $m^{*}$ is the cut-off time after which the Bell pair is discarded. 
%Here, we assume that all the link-level Bell pairs decohere similarly. 
%The clients $A$ and $B$ periodically attempt to generate link-level Bell pairs with a rate $R_{\textrm{clock}} = \frac{1}{T_{\textrm{clock}}}$. 
%Let $p_{A}$ and $p_{B}$ be the probabilities of successful link-level Bell pair generation corresponding to the users $A$ and $B$, respectively. 
%Therefore, the link-level Bell pair arrival rates are given by
%\begin{equation}
%    \mu_{A(B)} = R_{\textrm{clock}}\times p_{A(B)},
%\end{equation}
%and the arrival is modeled to be a Poisson process. 

%To describe our model, we use Poisson distribution as it represents the likelihood that a certain random variable $X$ number of events will occur within a predetermined time slot, provided the average number of times an event occurs during a given time slot is $\mu$, given by,
% \begin{equation}
%     P_{\textrm{Poisson}}(X = k) = \frac{\mu^{k}e^{-\mu}}{k!}
% \end{equation}

\section{Formulating Operations of the Quantum Switch as a Decision Process}
\label{QS_MDP}
A Markov decision process (MDP) is a stochastic process that describes a generic framework for sequential decision-making with random outcomes. 
In order to optimize a reward metric, a decision maker or agent continually interacts with the environment. The decision maker chooses a course of action that alters the state of the environment and may also yield a reward at the same time. The state evolution and reward may be random functions of the current state and chosen action. MDPs have been employed throughout a wide range of fields, such as flow control, artificial intelligence, management, manufacturing, automated control, and several others~\cite{Puterman2014-fz}. 
In the context of near-term quantum networking, an MDP was recently employed to determine policies for optimal elementary entanglement generation~\cite{khatri2022design}. 
In this work, we use the MDP formulation to identify optimal policies for distributing end-to-end Bell pairs by connecting randomly generated elementary Bell pairs in a quantum switch having two clients, with and without the option of entanglement distillation. 
In particular, we derive policies that optimize the end-to-end throughput of Bell pairs that exceed any given entanglement fidelity threshold.

% involving elementary Bell pairs of either clients as one of the available operations.

Our MDP requires the description of the following:
\begin{itemize}
    \item A set of possible MDP states:
    The state space of our MDP formulation of the quantum switch most generally consists of combinations (tensor products) of $N-$dimensional age vectors associated with $N\in Z^{\geq}$ elementary Bell pairs generated by client $A$ with the quantum switch and $M-$dimensional age vectors associated with $M\in Z^{\geq}$ elementary Bell pairs of client $B$, where these vectors are of the form  $[m_{1},m_{2},m_{3},\ldots] $, with $m_{i} \in (0,1,2....,m^{*})$ being the age of the $i^{th}$ elementary Bell pair. 
    We assume that all elementary Bell pairs have same cut-off age, i.e., $m^{*}_{i} = m^{*}$. 
   % We will refer to $i^{th}$ elementary Bell pair of client $A$ as $A_{i}$ and $k^{th}$ elementary link of $B$ as $B_{k}$. 
    %In our model, the elementary link associated with client B is immediately absorbed to create end-to-end entanglement with A. 
    We denote the state vector at any time $t$ as $S_{t}=[[m_{1},m_{2},m_{3},\ldots,m_N]_A,[m_{1},m_{2},m_{3},\ldots,m_M]_B]$, and the set of MDP states by $\mathcal{S}$.
    %= [S_{A},S_{B}]$ where $S_{A}\subset\mathcal{A}$ and $S_{B}\subset\mathcal{B}$.

        \item Given an MDP state, the switch performs actions based on a policy. 
        The list of possible actions that could feature in a policy of a quantum switch enabled with entanglement distillation are $\mathcal{A}$ = $\{a_{1},a_{2},a_{3}\}$:
    \begin{enumerate}[label=(\roman*)]
    \item $a_{1}$ is the action in which the switch decides to wait out the time-step in order for more link-level Bell pairs to be generated. 
    This is equivalent to no action being performed.
        \item $a_{2} $ is the action in which entanglement swap operation of elementary Bell pairs of the two clients are performed. 
        This action if successful, generates end-to-end entanglement between clients $A$ and $B$. If multiple swaps are possible this action also entails choosing which pairs to attempt to swap.
        \item $a_{3} $ describes the action of entanglement distillation of two elementary Bell pairs. 
        If and when successful, this action leads to creation of an elementary Bell pair of improved fidelity. 
        This action is performed to counter the effect of decoherence at the cost of sacrificing some elementary Bell pairs. If multiple distillations are possible this action also entails choosing which pairs to attempt to distill.
    \end{enumerate}
     \item $\mathcal{P}$ is a state transition probability matrix, where $P^{a}_{ss^{\prime}} = P[S_{t+1} = s^{\prime}| S_{t} = s, A_{t} = a] $ is the probability of transitioning from state $s$ to $s^{\prime}$ when an action $a\in \mathcal{A}$ is applied on the initial state and probabilistic link level entanglement is attempted:  
     In a unit time step, first probabilistic link-level entanglement with each client occurs; either or both may succeed or fail. Then the switch action is taken.
     % In effect, the state transitions from $s \rightarrow s^{\prime}$ and depending on probabilistic elementary entanglement generation, the state further makes a transition from $s^{\prime} \rightarrow s^{\prime\prime}$.

\item  A real-valued reward function, $r : \mathcal{S} \times \mathcal{A} \rightarrow \mathbb{R}$, where $r(s,a)$ is the reward obtained when the state transitions from $s\rightarrow s^{\prime}, \ \{s,s'\}\in\mathcal{S}$ when an action $a\in\mathcal{A}$ is applied. 
%The $S$ set contains any two states of $A$, $B$ or one of $A$ and one of $B$. 
If the state is `$s$' and the agent performs an action `$a$', we denote $r_{a}(s)$ as the immediate reward. 
Note that in our model, probabilistic link generation is always attempted and is not influenced by the switch's actions or decisions. Also, $r_{a}(s)$ could be an expectation, in the event of probabilistic rewards.
% Therefore, the state transition resulting from link generation is not associated with any reward.
   \begin{enumerate}[label=(\roman*)]
        \item The reward function for $a_{1}$ (No action), $r_{a_{1}}(s)$ is zero.
        \item The reward function for $a_{2}$, i.e.,
        entanglement swap between two elementary Bell pairs, say between the $i^{\textrm{th}}$ Bell pair with client $A$ denoted by $A_{i}$, and the $k^{\textrm{th}}$ Bell pair with client $B$ denoted by $B_{k}$, of fidelities $F_i$ and $F_k$, respectively, resulting in end-to-end entanglement is,
    \begin{equation}
        r_{a_{2}}(s) = q\times F_{ik} = 
         \begin{cases}
      0, & \text{if}\ F_{ik} < F_{th} \\
      1, & \text{otherwise}
    \end{cases}
    \end{equation}
    where, $q$ is the success probability of the entanglement swap operation, and $F_{ik}$ is the expression in (\ref{Werner_swap}) with $F_1=F_i$, $F_2=F_k$ and $F_{th}$ is the fidelity threshold for the end to end entanglement link.
    % \textbf{ We analyze two cases: one with a step function as the reward and the other with actual output fidelity values.}
    
\item The reward function for $a_{3}$, $r_{a_{3}}(s)$ is zero as  our metric is the creation of end-to-end entanglement links using entanglement swaps. While distillation alone does not create these links, it can move to a state which is more likely to perform entanglement swaps in future. This preparation is evaluated through the MDP.

% entanglement distillation between $A_{i}$ and $A_{j}$ that when successful, which happens with a probability $p_{\textrm{succ}}$ of (\ref{Werner_distill_p}), and results in a Bell pair of fidelity $F_{ij}$ of (\ref{Werner_distill_F}),
% is given by
% \begin{equation}
%    r_{a_{3}}(s) =  p_{\textrm{succ}}\times F_{ij}
% \end{equation}
\end{enumerate}

\item Discounted Reward:
The discount factor '$\gamma$' ranges from [0, 1], and scales down the future rewards over time such that the sum of rewards remains bounded.  The value zero indicates that the prospective rewards are completely irrelevant to us and the value one indicates that rewards in the future time step do not scale down. The parameter $H$ known as time horizon, describes a specific future time point at which certain processes will be assessed or presumed to come to a finish. It is a positive integer in the finite-horizon case or infinity in the infinite-horizon case.
A trajectory ($\tau$) is a collection of all the states and actions observed during an MDP process.
\begin{equation}
    \tau = \{s_{0},\mathfrak{a}_{0};s_{1},\mathfrak{a}_{1};\ldots,s_{H-1},\mathfrak{a}_{H-1};s_{H}\},
\end{equation}
where, $\mathfrak{a}_{i} \in \mathcal{A}$ is action applied by the agent at the $i^{th}$ time step on the state $s_{i}$.

The discounted sum of rewards is known as the return represented by $G_{\tau}$ in MDP,
\begin{equation}
    G_{\tau} = r_{0} + \gamma r_{1}+ \gamma^{2} r_{2}+ \gamma^{3} r_{3}+.....+ \gamma^{H}r_{H}
\end{equation}

\end{itemize}

To derive an optimal switching policy based within our Markov Decision Process (MDP), we have to make some assumptions. 
Firstly, we assume that the probability of success of the entanglement swap ($q$)  operation is one. 
This assumption is based on the fact that when Bell measurements are performed locally on matter-based quantum memories like trapped ion qubits~\cite{Duan2010-ul} or NV centers in diamond qubits~\cite{Yan2022-tq,Yin2015-ae}, it can  succeed with unit probability. Secondly, we consider only two clients with identical link generation success probabilities in a star topology.  
Thirdly, the quantum switch has a memory capacity that allows it to store up to a maximum of $L$ elementary Bell pairs for each client. 
Fourthly, we assume that the action taken by the switch and a probabilistic link generated are completed within each unit time step. 
Fifthly, in order to make our state space tractable, we consider only a discrete set of states in the MDP state space, which correspond to the entanglement fidelities associated with the possible ages of an elementary Bell pair $\{0,1,2,\ldots,m^*\}$. 
Associating states with all possible exact fidelities that arise from output of actions (in particular entanglement distillation) at the switch would lead to a rapidly increasing state space, which would make solving the MDP computationally infeasible. 
This assumption, however, necessitates mapping back the output of the entanglement distillation actions to states within the MDP state space by approximation: 
We map the fidelity of the newly distilled Bell pair to an age from the state space that has the closest fidelity. 
This assumption, thus, is built into the MDP and has an effect on identifying the optimal policy for the switch. Thus our policy is an approximation, or heuristic. To validate the results we perform simulations of the switch using the actual fidelity values of Bell pairs, and hence the results we report next (in terms of throughput, average fidelity, and jitter) are all valid, achievable values.

 % ......In order to find an optimal switching policy based on MDP tractable, we make few assumptions. These are,
 % \begin{itemize}
 % \item $q$ is the success probability of the entanglement swap operation which is $1$ for our switch as ..
 %     \item We consider only two identical clients in a star topology [cite].
 %     \item In a unit time step, action by the switch and probabilistic link generated is completed.
 %     \item We consider only discrete values in our state space represented ages of the elementary links. A more general description would require keeping track of both ages and  fidelities of the elementary links as we may have same ages of elementary links with two different fidelities. Such a treatment would make our state space very large making it intractable to solve. 
 %     \item As a consequence of previous assumption, we need to modify quantum distillation within physical considerations. Entanglement distillation can give rise states which do not belong to our state space, so we map the fidelity of newly distilled link to an age from the state space which has the closest fidelity but has a value which is lower than the newly distilled link. This assumption is within physical considerations as in an experimental setting, the quantum state will interact with the surrounding environment leading to a reduced value of fidelity. **This assumption is similar to the phenomenon of lattice re-normalization in which nodes are moved to bring back....**. 
 % \end{itemize}

\section{Optimal Policy and Results} \label{Results}

A policy is the decision-making process that guides a course of action. 
It is defined as a mapping from the state space to a distribution over the action space. 
A deterministic policy maps a state to a particular action when in that state. 
Mathematically speaking, it is 
\begin{equation}
\pi : \mathcal{S}  \rightarrow \mathcal{A}
\end{equation}
Thus, for any state $ s_{i} \in \mathcal{S}$, the agent takes an action $a_{j}\in\mathcal{A}$ according to the policy. 
For our switch model, we use policy iteration with discount factor $\gamma = 0.9$ to find the optimal solution~\cite{Puterman2014-fz}.

 \begin{figure}[htp]
       \centering
 \includegraphics[width=0.9\columnwidth]{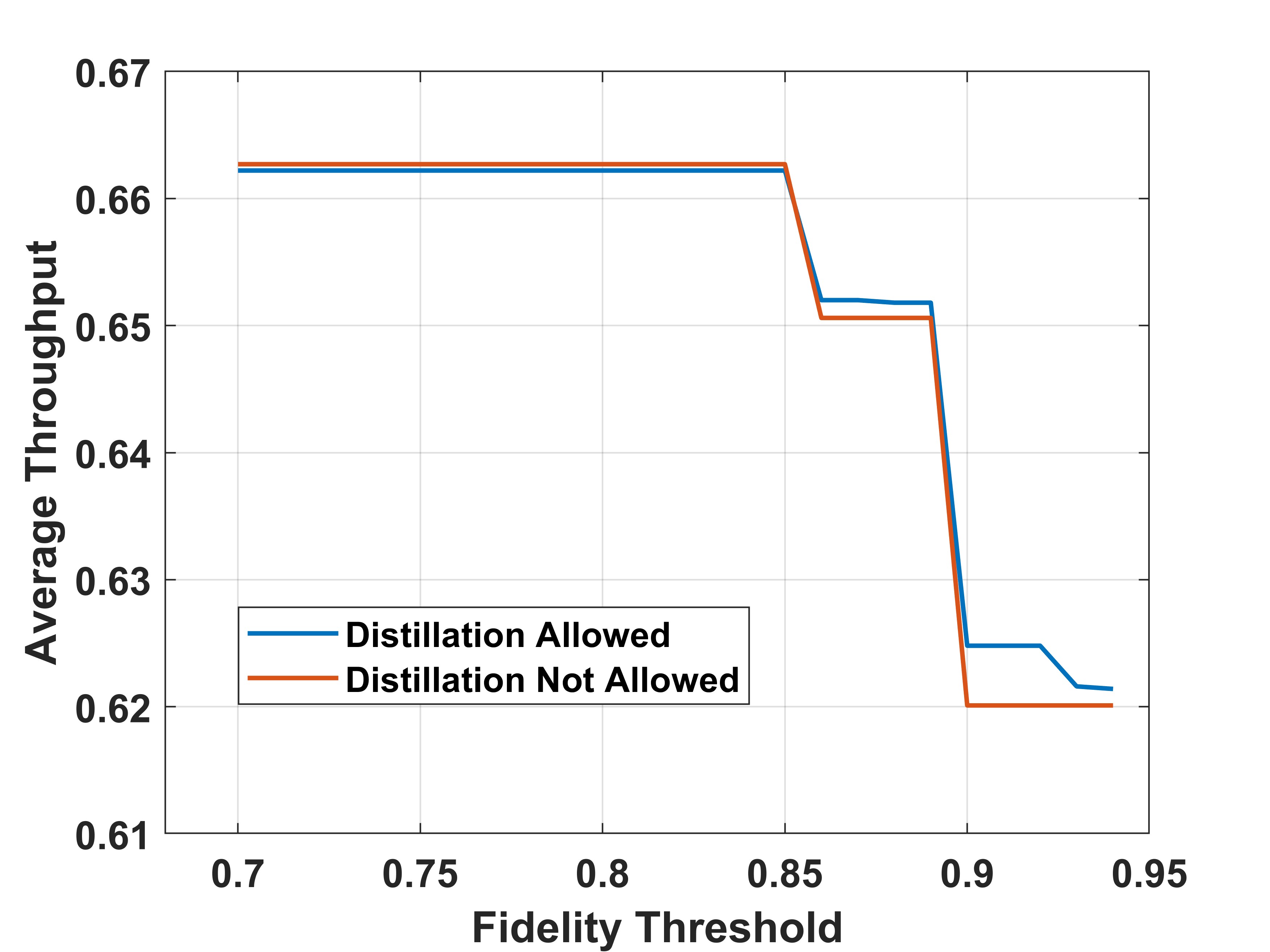}
        \caption{Average Throughput vs Fidelity Threshold for cutoff age $m^{*} = 3$, link generation probabilities, $\lambda_{1} = \lambda_{2} = 0.7$, cutoff fidelity $F^{*} = 0.85$ and  storage capacity of quantum memory, $L = 3$ for policies with and without entanglement distillation that optimize the end-to-end entanglement throughput.}
        \label{fig:AvgThru}
    \end{figure}

 \begin{figure}[htp]
       \centering
 \includegraphics[width=0.9\columnwidth]{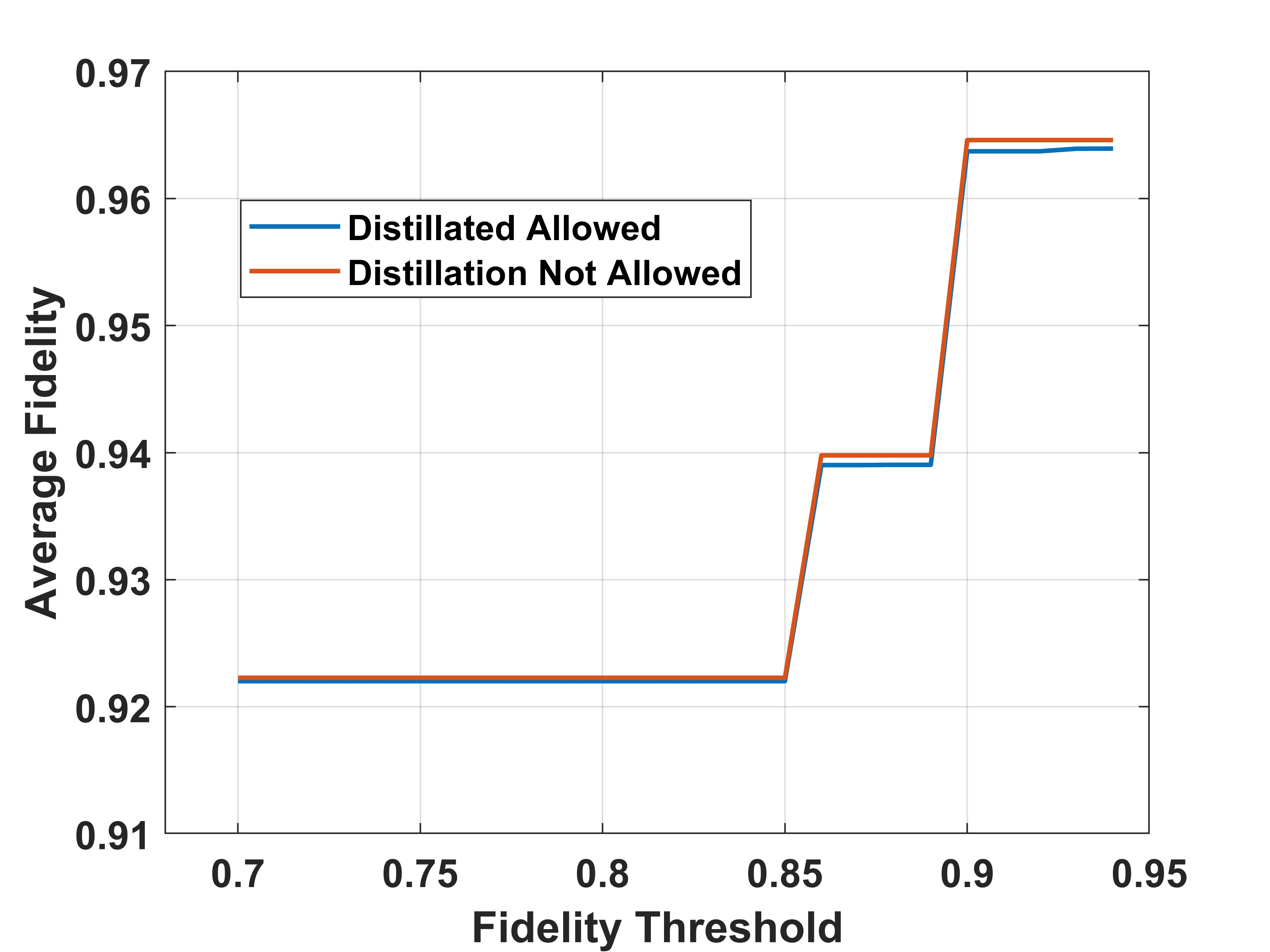}
        \caption{Average Fidelity vs Fidelity Threshold for cutoff age $m^{*} = 3$, link generation probabilities $\lambda_{1} = \lambda_{2} = 0.7$, cutoff fidelity $F^{*} = 0.85$ and  storage capacity of quantum memory, $L = 3$ for policies with and without entanglement distillation that optimize the end-to-end entanglement throughput.}
        \label{fig:AvgFid}
    \end{figure}

 \begin{figure}[htp]
       \centering
 \includegraphics[width=0.9\columnwidth]{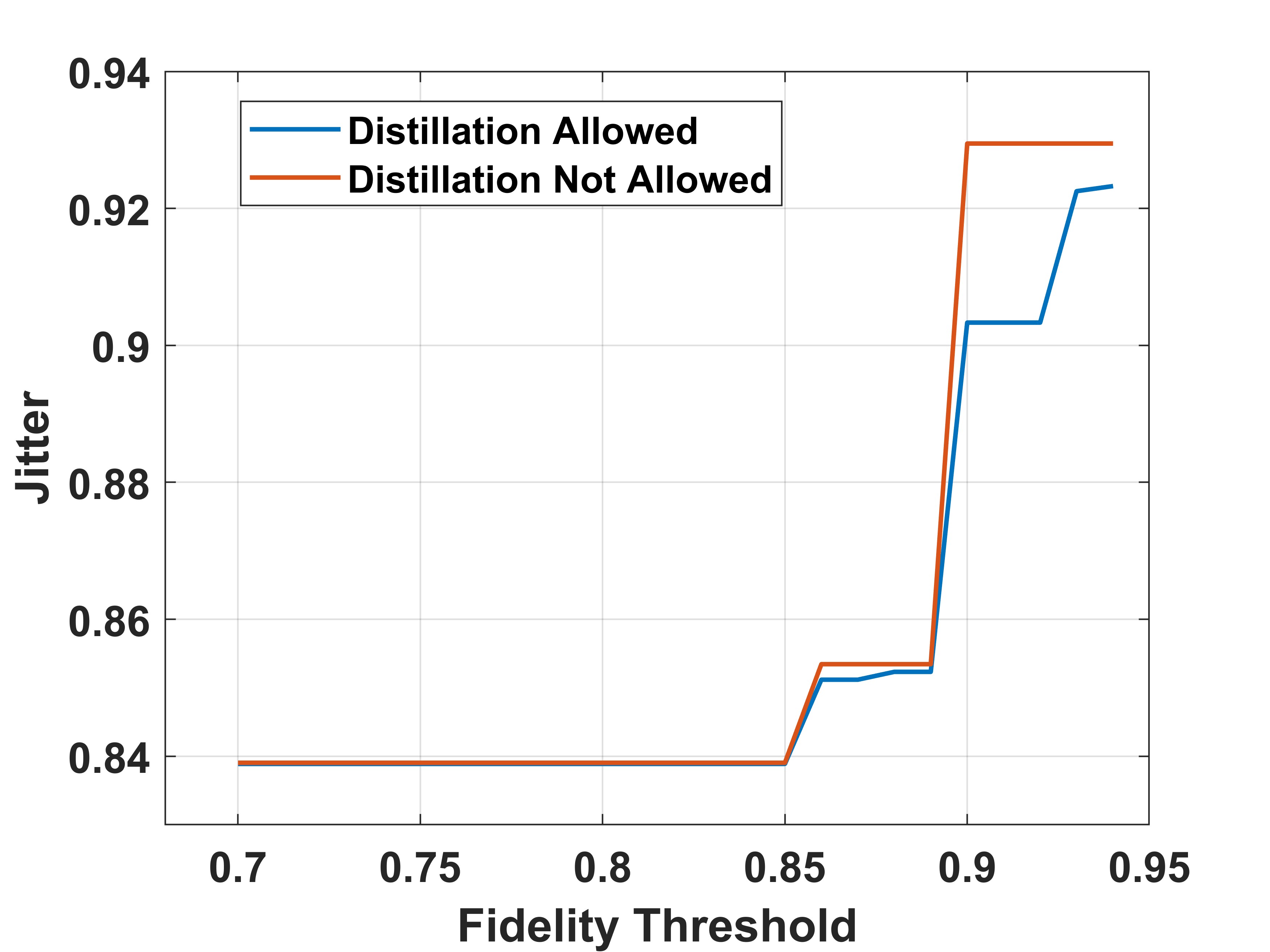}
        \caption{Jitter vs Fidelity Threshold for cutoff age $m^{*} = 3$, link generation probabilities $\lambda_{1} = \lambda_{2} = 0.7$, cutoff fidelity $F^{*} = 0.85$ and  storage capacity of quantum memory, $L = 3$ for policies with and without entanglement distillation that optimize the end-to-end entanglement throughput.}
        \label{fig:Jitter}
    \end{figure}

We solved the MDP for the quantum switch with and without entanglement distillation as an available action with the throughput of end-to-end entanglement above a given threshold fidelity as our objective function. 
We then simulated the switch's operation using the corresponding optimal policies for $10,000$ time steps. 
The results are plotted for three key metrics: average throughput, average fidelity, and jitter against fidelity which varies from 0.7 to 0.95.  Average throughput represents the number of end-to-end entanglement links above the fidelity threshold; average fidelity represents the average quality of these end-to-end entanglement links. Finally, jitter is the standard deviation of the time between successive end-to-end entanglements across all time steps.

Fig.~\ref{fig:AvgThru} shows the average throughput versus the fidelity threshold for the model with distillation (blue curve) and without (orange curve). 
At lower fidelities, the policy obtained through MDP that allows for distillation vs the one that does not, i.e., only based on entanglement swaps, yield the same average throughput. 
This is because entanglement swaps of low fidelity (decohered) elementary links can still generate an end-to-end link that meets the low required fidelity threshold. 
As the fidelity threshold increases, the optimal policy with distillation outperforms the policy that doesn't allow distillation, as not every pair of swaps can produce an end-to-end entanglement exceeding the threshold. 
% Both policies exhibit an average throughput of approximately $0.49$ beyond the fidelity threshold of $0.95$.
The reason behind this is that only entanglement swaps between the established highest fidelity (i.e. age 0) links can lead to end-to-end entanglements surpassing the fidelity threshold, thereby providing no benefit for distillation. Thus at intermediate target thresholds distillation, while it does not yield any end to end entanglement links itself, it enables the system to preserve fidelity of aging links, which yields greater throughput in the future.

The graph in Fig.~\ref{fig:AvgFid} shows the relationship between average fidelity and fidelity threshold for two different policies. The policy that allows distillation is optimized to achieve higher throughput rather than fidelity.
As a result, it tends to produce more end-to-end entanglements, as shown in Figure~\ref{fig:AvgThru}. However, this policy does not take into account the quality of the end-to-end links, as long as they meet the fidelity threshold. On the other hand, the policy that does not allow distillation always performs entanglement swaps to ensure high-quality links. Consequently, we observe that the average fidelity of the policy that does not distill is higher than the one where distillation is allowed.

The plotted graph in Fig.~\ref{fig:Jitter} shows the relationship between jitter and fidelity threshold. It reinforces the message that following an optimal policy which allows for distillation is superior to the one without distillation as it leads to a slightly lower interarrival time. Therefore, it can be inferred that the use of distillation in the optimal policy results in improved performance compared to policies that do not allow for distillation.

The plots of the three metrics indicate that the policy with quantum distillation performs better as compared to the policy with no distillation over a certain range of fidelity thresholds. However, beyond a threshold of 0.95, both policies have same values across all three metrics. This is because the policy with quantum distillation provides no advantage at such high fidelity thresholds.

\section{Conclusion and Future Directions}\label{Future_Direc}
This work provides evidence of the benefits of distillation in mitigating the effects of link decoherence in networks with minimum fidelity thresholds. 
Even though the magnitude of the gains are small in our examples, our model provides a foundation for further investigation into a wide range of scenarios. 
Our current model assumes symmetric link generation probabilities and queue lengths. 
We are currently exploring scenarios where these parameters differ across users. 
We have focused on the simplest case of two-to-one distillation. 
However, more complex protocols could be considered, where a higher $n$ number of elementary Bell pairs are transformed into a higher fidelity but fewer $k$ ($k<n$) number of elementary Bell pairs~\cite{Goodenough2023-nu, Rengaswamy2022-zz,Rengaswamy2021-fc,Krastanov2019-lj,Filip2018}. 
Our model here considers scenarios with only two clients. 
We are currently investigating scenarios where the switch must generate end-to-end bipartite entanglement for three or more users as well as multipartite entanglement. 
The existing model makes use of a discrete MDP state space, but in order to investigate more widespread situations, a more fine-grained state space that takes into account more possible entanglement fidelity values for elementary Bell pairs will be required. Reinforcement learning can be an alternative method to deal with state spaces that are high-dimensional and continuous. This is possible by utilizing neural networks to estimate the optimal policy, which determines the best action to take in a particular state. 
Further research in these areas will advance the study of quantum switches and will provide valuable insights into designing and optimizing them for increasingly practically relevant scenarios in quantum communication networks. 

\bibliographystyle{IEEEtran}
\bibliography{references.bib}

\end{document}